# Adaptive Item Calibration Via A Sequential Two-Stage Design

Yuan-chin Ivan Chang

Institute of Statistical Science, Academia Sinica, Taipei, Taiwan

**Abstract**

In this paper, we apply a sequential two-stage adaptive design to item calibration problems under a three-parameter logistic model assumption. In our procedure, the measurement errors of the estimates of the latent trait levels of examinees are considered. Moreover, a sequential estimation procedure is employed to guarantee that the estimates of the parameters reach a prescribed accuracy criterion when the iteration is stopped. Statistical properties of both the item parameter estimates and the sequential procedure are discussed. The performance of the proposed method is compared with that of the procedures based on some conventional designs using numerical studies.

# Adaptive Item Calibration Via A Sequential Two-Stage Design

**Introduction**

Due to the speed at which items are exhausted within computerized adaptive testing, the item calibration process is essential for this modern testing method. An efficient estimate of the unknown item parameters is demanded to improve item replenishment rate for such a modern computerized adaptive testing scheme. There are many different calibration processes being proposed and discussed under different scenarios, which can be found, for example, in Van der Linden and Glas (2000) and Wainer (2000), among others. In particular, Jones and Jin (1994) and Chang and Lu (2010) studied the calibration problem for two-parameter logistic (2PL) models using some measurement error model methods and D-optimal design schemes. Kingsbury (2009) and Makransky (2009) discussed the potential of adaptive and online item calibration processes under 2PL models. However, from the nature of a three-parameter logistic (3PL) model, the required optimal designs for estimating the guessing parameter are very different from those for the difficulty and discriminant parameters. Hence, directly applying those methods, which are mainly founded on 2PL models, to 3PL models cannot be efficient. This design issue makes the item calibration problem of a 3PL model different from that of a 2PL model.

Moreover, it is known from the experimental design literature that, in general, a sequential or a multiple-stage approach is required for finding the D-optimal for a nonlinear model (e.g., Kalish and Rosenberger, 1978 and Silver, 1980). Thus, due to the nonlinearity of 3PL models, the optimal designs must be constructed adaptively and sequentially; that is, the examinees selected to join the calibration process at the current stage will depend on the information obtained from all previous stages. So, the number of the examinees participated this procedure is sequentially increased and not predetermined, which is different from the conventional approaches that are based on a pre-fixed size of examinees such as in Ban, Hanson Wang, Yi and Harris (2001) and the references therein. Moreover, with adaptive sequential design methods, we are able to select the suitable examinees with latent trait levels that "match" parameters of items to be calibrated (Kingsbury, 2009), which will usually be computationally efficient in terms of number of examinees used in a calibration process. Furthermore, Patton, et al. (2013) studied the influence of the item calibration errors to adaptive testing. Hence, to fully take the advantage of the sequential design, a sequential estimation method is employed in order to control the precision of estimates. For a pre-specified precision of the parameter estimates, we propose a stopping rule that controls the number of examinees used in a calibration procedure. The asymptotic properties of the parameter estimates under such a stopping criterion are studied.

In the rest of this paper, we first discuss the design issues for item calibration under a 3PL model and the proposed two-stage design method. The asymptotic properties of the parameter estimates and sequential procedure are presented after that. We then evaluate the performance of the proposed procedure using synthesized data. Finally, this paper ends with a discussion. The highlights of proofs are given in the appendix.

## Sequential Design and Estimation

**Design Issues of Three-Parameter Logistic Models**

Let $\theta$ be the latent trait level of an examinee and Y be a binary response variable, where $Y = 1$ or 0 denotes whether the answer to a given item is correct or incorrect, respectively. Then a commonly used 3PL model in the item response theory (Wainer, 2000) is defined as

$$P(Y = 1 | \theta, a, b, c) = c + (1 - c) \{1 + exp[-a(\theta - b)]\}^{-1}, \qquad (1)$$

where $a$, $b$ and $c$ are the discriminant, difficulty and guessing parameter values of the item. For a given item, Equation 1 describes the probability of getting a correct answer from examinees with varying latent trait level $\theta$'s and the plot of probability of as a function of $\theta$ is called the item characteristic curve (ICC). If $c = 0$, Equation 1 becomes a 2PL model.

In a calibration process, we assign an item with unknown parameters to be calibrated to the selected examinees, and these item parameters are then estimated based on examinees' latent trait levels and responses to this item. The role of the selected examinees in a calibration process is similar to that of the design variables in a regression model (Silvey, 1980). However, unlike the conventional regression model, the different item parameters of a 3PL model require different design points, which can be easily illustrated with the figures below, and the geometric correspondence of the item parameters with an ICC.

Figure 1 shows the ICCs of items with fixed $b = 0, c = 0.1$ and different values of $a$. Note that all these ICC curves have the same asymptote as $\theta$ goes to $-\infty$, which only depends on the value of $c$. That is, estimating the guessing parameter $c$ of an item is like estimating the intercept of the asymptote of its ICC to the probability axis. Hence, for this purpose, we mainly need examinees with low latent trait levels. In addition, from this figure we also find that, for a fixed difficulty level $b$, the ICC of an item with a small $a$ decreases more slowly than for those items with larger values of $a$. This situation makes the estimate of $c$ for an item with a small value of $a$ more difficult, since we will need to locate examinees with even lower latent levels, which are rare in practice. Likewise, it also implies that to estimate $c$ for an item with low difficulty level $b$ can also be difficult even when its $a$ value is large.

[Insert Figure 1 here.]

The parameters $a$ and $b$ also have their own geometric meanings regarding the ICC. Parameter $b$ is a location parameter of ICC. Thus, to estimate $b$, we can simply put designs around two sides of $b$. The value of $a$ equals the slope of the tangent line going through the inflection point $(b, 0.5 + 0.5c)$ of the ICC in the $\theta$-probability coordinate shown in Figure 2 (a). That is, estimating $a$ is equivalent to estimating the slope of this tangent line. Thus to estimate an item with a small value of $a$, we should select examinees with $\theta$'s in a wider range around $b$ than for an item with a large $a$. On the contrary, if $a$ is large, then the tangent line becomes very steep. Thus in order to have a good estimate of $a$ in this situation, we should locate $\theta$ within a narrow neighborhood of $b$. In fact, the responses from examinees with $\theta$'s far away from $b$ cannot provide much information about the slope of this tangent line, but add some noise to the estimate of $a$ (Figure 2 (a)). This situation is true for all items, but especially in the case of an item with large $a$.

[Insert Figure 2 here.]

These kinds of geometric viewpoints from ICCs are further supported with plots of information curves. Figure 3 (a) shows plots of the determinant of the information matrices of different $a$'s as a function of $\theta$ with fixed $b = 0$ and $c = 0.1$. Their curves are very similar to those of the 2PL models and have only two peaks of curves slightly shifting to the right due to the parameter $c$. That is the reason why we consider putting these two parameters in one group and applying a D-optimal design method for estimating them as in the 2PL model case. From the information curves of $c$, as functions of $\theta$ for different pairs of $(a, b)$, it is clearly shown in Figure 3 (b) that the smaller the value of $\theta$, the higher the information curve of $c$. That is, in theory, the maximum information for estimating $c$ is achieved as $\theta$ goes to negative infinity. However, it is impossible to have examinees with latent trait levels $\theta$ close to $-\infty$. So, we can only find examinees with relatively lower latent trait levels, which is item-dependent for this purpose. Moreover, the range of $\theta$'s for estimating $c$ will also depend on parameters $a$ and $b$. We can see from Figure 3 (b) that the information curves of $c$ for items with large $a$ reach their maximums much faster than items with small $a$. Contrary to the situation of $c$, the maximum information of $a$ and $b$ together occurs around $b$, and the distance between the two information peaks becomes smaller as $a$ becomes larger such that the maximum information occurs in an even narrower neighborhood of b. Therefore, when we apply a D-optimal design method directly to a 3PL model, the amount of information for individual parameters may be diluted, because this method will treat the different needs of three parameters equally. This suggests that directly applying a strict D-optimal design scheme to the calibration problem of a 3PL model cannot be as efficient as expected, since it is difficult to have design points that satisfy these different needs at the same time.

[Insert Figure 3 here.]

**Selecting Examinees For Estimating Guessing Parameter $c$**

We know that $c$ is the probability-axis intercept of the asymptote of an ICC as $\theta$ goes to $-\infty$ and the ICC is "reflectional symmetric" about the reflection point $(b, (1 + c)/2)$ in this $\theta$-probability coordinate (Figure 2 (a)). Moreover, it is easy to see that the line *probability* $= 1$ is another asymptote of ICC as $\theta$ goes to $+\infty$. Equation 1 implies that for given parameters $a$, $b$ and $c$, the ICC is strictly increasing in $\theta$, thus for a given probability $p_0$ near 1, there is a unique $\theta_h$ such that $ICC(\theta_h) = p_0$. It implies that

$$\theta_h = b + a^{-1} \log (p_0 - c)/(1 - p_0 ).$$

Let

$$\theta_\ell = 2b - \theta_h = \mathrm{b} - a^{-1} \log (p_0 - c)/(1 - p_0 ), \tag{2}$$

then it follows that $ICC(\theta_\ell) = c + (1 - p_0)$. Hence, if $p_0$ is close to 1, then the probability for examinees with latent trait levels less than $\theta_\ell$ to answer such an item correctly is close to $c$ (Figure 2 (b)). This suggests that, if estimates of $(a, b)$ are available, then we can determine the range of $\theta$ for estimating $c$ by choosing a suitable $p_0$ close to 1.

**Selecting Examinees for Estimating Parameters $a$ and $b$**

Suppose $c$ is a known constant in a 3PL model and let

$p' = (p-c)/(1-c) = \{1+\exp[-a(\theta-b)]\}^{-1}$, where $p = P(Y=1|\theta,a,b,c)$. It implies that

$\log[p'/(1-p')] = a(\theta-b)$, which is a 2PL model. It suggests that if $c$ is fixed, then we may still apply a D-optimal design scheme to find suitable examinees for estimating $a$ and $b$ as in 2PL models. For an ordinary logistic model, Kalish and Rosenberger (1978) claimed that the D-optimal design for a logistic regression model should put equal design points at $L_{17.6}$ and $L_{82.4}$, where $L_p$ denotes the $p$-th quantile of a logistic distribution. Hence, following the suggestions of Kalish and Rosenberger (1978), we should choose equal sample sizes of examinees with $\theta$ satisfying $ICC(\theta) = c + (1-c)L_p$, for $p = 17.6$ and $82.4$. (Note that here the $L_p$ are still quantiles of an ordinary logistic distribution, and are not from a 3PL model.)

Thus, we need to estimate the unknown parameter $c$ first, before we can construct the designs for $(a, b)$ and a good estimate of $c$ will certainly assist with finding the best designs for $(a, b)$. On the other hand, from the previous discussion, to find suitable examinees for estimating $c$, we also need the information of $(a, b)$. That is the reason why an iterative algorithm, which allows us to update the information regarding the unknown parameters and separately select designs for estimating $(a, b)$ and $c$, is required.

**Two-Stage Iterative Design Procedure**

Following above discussions, we propose a two-stage approach such that the best designs for estimating $c$ and $(a, b)$ can be separately selected at each stage. We then apply this two-stage design iteratively, and based on the updated information about these parameters, examinees are sequentially selected for joining the calibration process. In addition, with a prefixed estimation accuracy for the item parameters, we proposed a stopping criterion on the iterative design procedure such that the estimates will fulfill the precision criterion when the calibration procedure of an item is stopped (Siegmund, 1985). A sequential iterative two-stage design scheme with a stopping criterion is stated as follows:

**Step (1) Initial estimate of $c$:** Randomly select a fixed number of examinees from examinees with estimated latent trait level $\theta < \theta_c$ for some prefixed $\theta_c$, and estimate the guessing parameters $c$ based on their responses; for example, the proportion of correct responses is sufficient to be used as an initial estimate of c in this step.

**Step (2) Initial estimates of $(a, b)$:** Randomly select a batch of examinees with latent trait levels $\theta \in \Theta$, where $\Theta$ denotes some prescribed range of $\theta$; say, for example, $\Theta = (-3.6, 3.6)$. Let $c$ equal to the estimate obtained in Step (1), and then calculate estimates of $(a, b)$ using all responses up to this stage.

**Step (3) Iterative step:** Using the estimates of $(a, b)$ obtained in Step (2), we then select a new batch of examinees with latent trait levels lower than $\theta_\ell$ as defined in Eq. (2). (That is, we use $\theta_\ell$ as a new upper bound of latent trait levels for choosing new examinees to update the estimate of c.) In addition, using the estimates of $(a, b)$ and $c$ from Steps (1) and (2) together, we can select another batch of examinees based on the D-optimal design scheme (Kalish and Rosenberger, 1978) and then re-estimate all three parameters *using all of the collected responses* and employing the maximum likelihood estimation method.

**Step (4) Stopping:** Repeat Step (3) until a prescribed stopping criterion of estimates is satisfied.

In Step (1), the initial choice of $\theta_c$ is not that crucial, since the range of design points for parameter *c* will be updated in the following iteration steps. Under such an iterative scheme, the updated information of the unknown parameters is integrated into the design method and the suitable examinees are selected separately at each stage. The stopping rule we used in Step (4) is based on a sequential confidence set estimation scheme, and the details are described in the following section. Because the Fisher information matrix is a complicated function of three unknown parameters, it is not easy to clarify the individual impact of each item parameter on the information matrix. Hence, there is no theoretical result for the optimal ratio of batch sizes. In our numerical study, the batch sizes ratio of (*a*, *b*) to *c* is equal to 2:1 during the iteration steps.

Note that, since new design points are determined based on previous estimates of parameters and all responses from the selected examinees are used to estimate the parameters, the independent observations assumption is no longer appropriate in this case. Hence, the method of stochastic regression with adaptive designs must be used. For more detailed discussion about the advantages and disadvantages of the item calibration using adaptive or random designs, please refer to Kingsbury (2009). Moreover, if the examinees are selected based on their estimated latent trait levels instead of the true levels, the measurement error methods (Carroll, Ruppert, Stefanski and Crainiceanu, 2006; and Jones and Jin, 1994) are naturally involved. In the following section, we will discuss the asymptotic properties of parameter estimates under such a measurement error, adaptive design scenario.

## Asymptotic Properties of Estimates of Item Parameters

**Estimation of Item Parameters Under Two-Stage Design**

For convenience, we follow the re-parameterization scheme of Jones and Jin (1994) by defining $\beta = (-ab, a)'$ such that the D-optimal design method for logistic regression models can be applied to $\beta$ directly and let $x = (1, \theta)'$. Then the 3PL model is re-written as:

$$P_i \equiv P(Y_i = 1|\beta, c) = c + (1-c)G_i(\beta) \text{ for } i = 1, 2 \ldots,$$

where $G_i(\beta) = (1 + exp(x'_i\beta))^{-1}$. So, the likelihood function of a 3PL model of a sample of size n is

$$L_n(\theta, \beta, c) = \prod_{i=1}^{n} P_i^{Y_i}(1-P_i)^{1-Y_i}.$$

Let $\gamma \equiv (\beta', c)'$ and $\ell_n \equiv \ell_n(\theta, \gamma) = ln\,[L_n(\theta, \beta, c)]$ be the log-likelihood. Assume that, for each i, instead of the true $\theta_i$, we only observe $\tilde{\theta}_i = \theta_i + \xi_i$, where $\tilde{\theta}_i$ denotes the observed latent trait level and $\xi_i$ is its corresponding measurement error. Let $\tilde{x}_i = (1, \tilde{\theta}_i)' = (1, \theta)' + (0, \xi_i)' = x_i + (0, \xi_i)'$ be the design with measurement error.

Then the maximum likelihood estimate (MLE) of $\gamma$ can be obtained by solving the following equation:

$$\nabla \ell_n(\gamma) \equiv (\partial \ell_n)/\partial \gamma = ((\partial \ell_n)/\partial \beta', (\partial \ell_n)/\partial c)' = U_n(\gamma, \tilde{x}, Y) = 0. \qquad (3)$$

By the Multivariate Mean-value Theorem,

$$\nabla \ell_n(\gamma) = \nabla \ell_n(\gamma_0) + \nabla^2 \ell_n(\gamma^*)(\gamma - \gamma_0) = 0,$$

where $\gamma_0 = (\beta'_0, c_0)'$ is a vector of true parameters, $\gamma^*$ is in the line segment of $\gamma$ and $\gamma_0$, and

$$-\nabla^2 \ell_n(\gamma^*) = - \begin{bmatrix} \frac{\partial^2 \ell_n}{\partial \beta^2} & \frac{\partial^2 \ell_n}{\partial \beta \partial c} \\ \frac{\partial^2 \ell_n}{\partial c \partial \beta} & \frac{\partial^2 \ell_n}{\partial c^2} \end{bmatrix}_{\gamma=\gamma^*}. \tag{4}$$

Assume further that all item parameters are bounded; that is, $a \in [m, M]$ for some $M > m > 0$, $b \in [B_m, B_M]$ for some $B_m, B_M \in R$, and $c \in [0,1]$. Let $\mathcal{J}_{\gamma^*} \equiv -\nabla^2 \ell_n(\gamma^*)$. Then with probability one, $\hat{\gamma}$ satisfies

$$\hat{\gamma} \approx \gamma_0 + \mathcal{J}_\gamma^{-1}(x) U_n(\gamma_0, x, Y) + O(n^{-1} \sum_{i=1}^n \xi_i^2). \tag{5}$$

Let $\mathcal{F}_n$ be a sequence of σ-fields with $\mathcal{F}_0 = \sigma\{\emptyset, \Omega\}$ satisfying $\mathcal{F}_n \subseteq \mathcal{F}_{n+1}$ for all $n \geq 1$ such that $x_i$ is $\mathcal{F}_{i-1}$-measurable and $Y_i$ is $\mathcal{F}_i$-measurable. Let $\varepsilon_i = Y_i - E[Y_i|\mathcal{F}_{i-1}]$ for all $i \geq 1$ be the error of the model. Then $\varepsilon_i$ is $\mathcal{F}_i$-measurable for all $i \geq 1$. By the definition above, we have $\tilde{x}_i = x_i + (0, \xi_i)' \in \mathcal{F}_{i-1}$, where measurement errors $\xi_i \in \mathcal{F}_{i-1}$. Let $\lambda_{min}(n)$ and $\lambda_{max}(n)$ be the minimum and maximum eigenvalues of $\sum_{i=1}^n x_i x_{i\prime}$, respectively. Assume further that

(C1) $\lambda_{min}(n) \to \infty$ a.s. and $\lim_{n\to\infty} \lambda_{min}(n)/\log(\lambda_{max}(n)) = \infty$ a.s.,

(C2) $\sup_{i\geq 1} E(|\varepsilon_i|^p|\mathcal{F}_{i-1}) < \infty$ a.s. for some $p > 2$,

(C3) there exists a sequence of non-random matrices $B_n$ such that $B_n^{-1}(\sum_{i=1}^n w_i w_{i\prime})^{\frac{1}{2}} \to_p I_3$ and $max_{1\leq i \leq n} \| B_n w_i \| \to_p 0$, where $w_i = (1, \theta_i, 1)' = (x_{i\prime}, 1)'$, for all $i \geq 1$, and $I_3$ is the 3×3 identity matrix, and

(D) $E[\xi_i|\mathcal{F}_{i-2}] = 0$ for all $i$ and there is an increasing sequence of positive constant $m_n$ such that $sup_{n\geq 1} E[(\sum_{i=1}^n |\xi_i|/\sqrt{m_i})^2] < \infty$ as $n \to \infty$ almost surely.

Then we have following asymptotic properties of $\hat{\gamma}$:

**Theorem 1.** *Suppose item parameters $a \in [m, M]$ with $0 < m < M$, $b$ is bounded and $c \in [0,1]$. Assume that $sup_{i\geq 1}|\theta_i| < \infty$, and Conditions (C1), (C2) and (D) are satisfied with $m_n = O(n)$,* then $\hat{\gamma}_n \to \gamma_0$ almost surely as n goes to infinity. If, in addition, Condition (C3) and (D) are satisfied with $m_n = O(n^{1/2})$, then $\hat{\mathcal{J}}_{\hat{\gamma}}^{1/2}(\hat{\gamma}_n - \gamma_0) \to_L N(0, I_3)$, where $\hat{\mathcal{J}}_{\hat{\gamma}} \equiv -\nabla^2 \ell_n(\hat{\gamma})$.

Note that Condition (C1) is a regularization condition on the design matrix, which requires that the design matrix is eventually positive definite with a rate approximately equal to *log(n)* when the sample size *n* becomes large. Condition (C2) is a conditional moment condition on the modeling errors. Condition (C3) is required for the martingale central limit theorem, since the independent assumption is no longer valid under an adaptive design scenario. Condition (C1) – (C3) were first discussed in Lai and Wei (1982), where they studied the linear stochastic regression. Because for a 3PL model, $\varepsilon_i$ is bounded and we also assume that both latent trait levels and parameter values are bounded, all these conditions are satisfied. Condition (D) is only required when the measurement error is considered (Chang, 2011), which will be satisfied if the variance of the estimates of the latent trait levels become smaller at a certain rate when the number of items used to estimate them becomes large. This condition is satisfied when a variable length computerized adaptive test is used. The sequence $m_n$ describes the decay rate of measurement error as sample size increases. This can usually be achieved by increasing the number of items used to estimate the latent trait level of an examinee under a variable length adaptive testing scheme (Chang and Lu, 2010). If $m_n$ goes to infinity more slow, then we need to have $\xi_n$ decay faster such that Condition (D) holds. The highlights of the proof of Theorem 1 are given in the Appendix.

**Estimation Accuracy and Stopping Rule**

From Theorem 1, we have $(\hat{\gamma}_n - \gamma_0)' \hat{\mathcal{J}}_{\gamma} (\hat{\gamma}_n - \gamma_0) \to_L \chi^2(3)$ as n → ∞, where $\chi^2(3)$ denotes the chi-square distribution with degrees of freedom equal to 3. Let

$$R_n = \{\gamma \in R^3 : (\hat{\gamma}_n - \gamma)'\hat{\mathcal{I}}_\gamma(\hat{\gamma}_n - \gamma) \leq C_\alpha^2\},$$

where $C_\alpha^2$ is a constant satisfying $P(\chi^2(3) \geq C_\alpha^2) = \alpha$. Then $R_n$ defines a confidence ellipsoid for $\gamma_0$ centered at $\hat{\gamma}_n$ with coverage probability approximately equal to $1 - \alpha$ as sample size $n$ becomes large. Suppose that the parameter $\gamma_0$ is known, then the optimal sample size for constructing $R_n$, with the length of its maximum axis no larger than 2d and coverage probability equal to $1 - \alpha$, is

$$n_{opt}(d) = \inf\left\{n \geq n_0 : n\lambda_{\gamma_0} \geq \frac{C_\alpha^2}{d^2}\right\}, \quad (6)$$

where $n_0$ denotes the initial sample size and $\lambda_{\gamma_0}$ is the minimum eigenvalue of

$$\mathcal{I}_0 \equiv -E\left\{\begin{bmatrix} \frac{\partial^2 \ell_1}{\partial \beta^2} & \frac{\partial^2 \ell_1}{\partial \beta \partial c} \\ \frac{\partial^2 \ell_1}{\partial c \partial \beta} & \frac{\partial^2 \ell_1}{\partial c^2} \end{bmatrix}_{\gamma=\gamma_0}\right\}.$$

That is, $n_{opt}(d)$ is the minimum sample size required for constructing such a confidence ellipsoid $R_n$ with the required properties, when all three parameters $(a, b, c)$ are known. Of course, this optimal sample size cannot be used in practice, since it depends on the true $\gamma_0$, which is the unknown vector of parameters to be estimated. However, it follows from the strong consistency of $\hat{\gamma}_n$ that we have $lim_{n\to\infty} n^{-1}\hat{\lambda}_{min}(n) = \lambda_{\gamma_0}$ almost surely, where $\hat{\lambda}_{min}(n)$ is the minimum eigenvalue of $\hat{\mathcal{I}}_{\hat{\gamma}}$. Hence, replacing $\lambda_{\gamma_0}$ in Equation 6 with $n^{-1}\hat{\lambda}_{min}(n)$, we have the following stopping criterion:

$$T_d = \inf\left\{n \geq n_0 : \hat{\lambda}_{min}(n) \geq \frac{C_\alpha^2}{d^2}\right\}. \quad (7)$$

The properties of the item parameters' estimates under the proposed stopping criterion are summarized below.

***Theorem 2.*** *Under the assumptions of Theorem 1, if Conditions (C1) and (C2) are satisfied and if there is a sequence of positive real number $m_n = O(n)$ such that Condition (D) is satisfied, then (i) $T_d < \infty$ for any $d > 0$, $T_d \to \infty$ as $d \to 0$, and $lim_{d \to 0} T_d/n_o = 1$ and $lim_{d \to 0} \hat{\gamma}_{T_d} = \gamma_0$ with probability one. Moreover, if Conditions (C3) and (D) are satisfied with $m_n = O(\sqrt{n})$, then (ii) $lim_{d \to 0} P(\hat{\gamma}_{T_d} \in R_{T_d}) = 1 - \alpha$.*

Theorem 2 says that under the proposed stopping criterion $T_d$, when the sampling is stopped, the estimate of $\gamma_0$ is strongly consistent and has the asymptotic normality property. In addition, the confidence set $R_{T_d}$ also has the nominated coverage probability, approximately, and the ratio of the sample size used based on this stopping criterion ($T_d$) to the best sample size ($n_{opt}$) converges to 1 almost surely as $d$ goes to 0.

## Numerical Examples

We focus on the estimation of item parameters and the efficiency of the procedure. We also assume that the latent trait levels of examinees selected for the calibration process are measured with errors. For other general issues, such as the test administration of methods using different designs, please refer to Kingsbury (2009) and the reference therein.

In our simulation studies, we take $a \in \{0.5, 1, 1.5, 2\}$, $b \in \{-2, -1, 0, 1, 2\}$ and $c = 0.1$. The true latent trait levels of examinees, $\theta$, are generated with in the range $(-3.6, 3.6)$. In addition, the measurement error $\xi_n$ is generated from a normal distribution with mean 0 and standard deviation equal to $0.5/\sqrt{n}(log(n))^{1.1}$ as that in Chang and Lu (2010). This is reasonable choice because the estimate of the latent trait level of an examinee can usually be proved to be asymptotically normally distributed; for example, see Chang and Ying (2004). So, Condition (D) is satisfied with $m_n = \sqrt{n}$. It is also easy to check that Conditions (C1)-(C3) are eventually satisfied unless all examinees selected have the same latent trait level, which is very unlikely to happen in practice.

In all simulation runs, the initial sample sizes for estimating $(a, b)$ and $c$ are 100 and 10, respectively, and $\theta_c = -2$. In Step (3), we then select 10 examinees for $(a, b)$ and 5 examinees for $c$ at each iteration according to their corresponding selection schemes, which are discussed earlier. For the procedure with a strict D-optimal design scheme, all the parameter setups are the same except for the design part, in which 15 examinees are selected according to the numerical solution, based on the collected observations up to the current design iteration stage, of the maximum point of the determinant of the information matrix. (In simulation studies, we fix $c = 0.1$ for convenience purposes. However, following the discussion in Section 2, it is easy to note that, when $c$ becomes larger, locating the examinees for estimating $c$ becomes easier, which will also result in improving the design selection of $(a, b)$. Thus, the required sample sizes may be reduced as $c$ becomes smaller than that of $c = 0.1$.) Note that in our numerical studies, the sizes of two batches of examinees selected for $(a, b)$ and $c$, during the iteration, are 10 and 5, respectively.

Table 1 summarizes the results of parameter estimates based on 1,000 runs using the proposed method with $d = 0.5$ and coverage probability equal to 0.95 (i.e. $\alpha = 0.05$). The averages of estimates for different parameter settings are recorded and their standard deviations are stated within parentheses. The last three columns of this table are the individual mean square errors (MSE) for three parameters. It can be seen from this table that the estimates for all three parameters have very small MSE. The estimates of $c$ are very stable in general, except for the cases of $a = 0.5, b = -2$ and $a = 0.5, b = -1$. As discussed before, to estimate $c$ in these two cases we need examinees with very small latent trait levels. However, in our simulation study, the $\theta$ are confined within a fixed range $\Theta = (-3.6, 3.6)$ and that is the reason why the MSEs of the estimates of c in these two cases are slightly larger than those of other cases.

Table 2 summarizes the results of the estimates based on a strict D-optimal design scheme under the same simulation setup used in the 2-stage design procedure. That is, here we skip Step (1) in the proposed algorithm and the optimal design is constructed with direct optimization of the determinant of information matrix using an initial sample size of 110 randomly selected from $\theta \in (-3.6, 3.6)$ as in Step (2). The stopping criterion is the same with $d = 0.5$ and $\alpha = 0.05$. Comparing the results in Table 2 to those in Table 1, we can see that, although the MSE's of the estimates obtained using a strict D-optimal design are still in a tolerable range for most of the cases, they are clearly larger than those of the estimates based on the proposed two-stage method. It is also worth noting that the estimates of $b$ in the cases with $a = 0.5$ and $b = -1$ and $-2$ are worse than in other cases. This is mainly caused by the quality of estimates of $c$ in the earlier stages for these cases, and this situation is consistent with our previous discussion.

Table 3 reports the average sample sizes (stopping times) and the individual coverage probabilities of three parameters of the two procedures based on 1,000 runs for each combination of parameters. Items with different values of $a$'s are divided into four groups by vertical dot-lines. It is clearly shown that, for fixed $c$, the larger the value of $a$, the more samples are required to estimate the item parameters. In addition, if we only compare the sample sizes used with the same $a$ value, then for most cases the larger the absolute values of $b$, the more samples are required such that the parameter estimates can achieve the same precision. We now compare the sample sizes used in these two procedures. We can see from this table that for most of the combinations of parameters, the proposed procedure is more efficient than the procedure based on a strict D-optimal design in terms of the average sample sizes used for constructing a confidence ellipsoid of parameters with the same fixed length of maximum axis and prescribed coverage probability. For some items with large values of $a$ and $b$, a strict D-optimal design procedure outperforms the proposed procedure. This is because when $a$ is large, the estimation of $c$ is relatively easy, and thus the choice of design points will not be affected much by the instability of the estimate of $c$. Therefore, the strict D-optimal design procedure can take full advantage of the D-optimal design in these cases. However, when both $a$ and $b$ are large; for example, $a = 2$ and $b = 2$, both procedures become unstable, which is partly due to the limited range of $\theta$'s used in our simulation study. In addition, it is also due to the value of $a$. When $a$ becomes large, the peaks of the determinant of information matrix (as a function of $\theta$) are confined to a smaller range such that the optimal design points are very difficult to locate. This is especially the case in procedures that are based on a strict D-optimal design. As shown in this table, both the average sample size and its standard deviation become very large for the case with $a = 2$ and $b = 2$, and we suspect that the difficulty in this case may also be enlarged by the estimate of $c$ under such a strict D-optimal design method.

    The coverage probabilities of both methods are larger than the target 95% for most of the item parameters, and this is especially the case for the two-stage design, which suggests that both methods are conservative and may over-sample the examinees selected to join the calibration. It is well known that the stopping criterion that controls the maximum axis of the confidence ellipsoid is conservative, which is one of the reasons that the coverage probabilities are higher than 95%. Some fine tuning of the procedure to reduce the over-sampling situation is possible; for example, we may dynamically change the sample sizes in Step (3) according to the estimating accuracy of parameters. However, this is beyond the scope of this study.

We also compare our results with those obtained using a random design scheme assuming that the latent trait levels follow a normal distribution with mean 0 and standard deviation 1.16 such that about 99% of the generated latent trait levels will fall into to (-3, 3) with the same measure errors as before. The examinees participating the calibration process are then randomly selected. In Table 4, the averages of estimates of parameters and their corresponding standard deviation based on 1000 runs are reported. The average number of examinees used for calibration of each item is also included. Because a stopping criterion is used, the coverage probabilities based on the random design are similar to the other two cases, and therefore are not included here. In general, due to the distribution of the latent trait levels that we sample from, the estimates of item with difficulty parameter $b$ close to 0 will usually perform better than that of other items. The items with small $a$ and $b$ are estimated poorly using random design (Table 4). This is because when the parameter values of ($a$, $b$) are small, it is very difficult to estimate parameter $c$ using a random design. When values of parameter $a$ and $b$ become larger, the precisions of estimates are clearly improved. This situation also explains why there is an advantage of using a two-stage design. However, although the precision of estimates of item is improved when the values of item parameter $a$ and $b$ become large, the method with a random design usually requires a larger number of examinees to achieve the targeted estimating accuracy and coverage probability than that of the proposed method. We found that the proposed method requires the smallest sample sizes to achieve the specific goal among the three methods.

**Discussion**

In this paper, we propose a sequential two-stage design scheme for 3PL online calibration processes of computerized adaptive testing. The sequential fixed-size confidence set estimation

20method is used to guarantee the precision of item parameter estimates and the measurement error in the latent trait level estimate is considered. We know that the performance of the D-optimal design method for a nonlinear model depends highly on the initial choice and the ongoing estimation of the unknown parameters. If we apply the strict D-optimal design method to a 3PL model, then the estimates of the guessing parameter $c$ in the early stages are much worse than the ones using only the examinees with low latent trait levels, and these "bad" estimates of $c$ result in bad designs in the following steps. The proposed method locates those examinees with low latent trait levels using the geometric properties of ICC for estimating the guessing parameter, which is numerically stable and efficient than the estimate obtained from a strict D-optimal design in the 3PL calibration problem. This estimate of $c$ provides a good beginning for the iterative design scheme, and the separate design selection method leads to a more efficient estimation of parameters $(a, b)$, which also helps to locate suitable examinees for estimating $c$ again in the following steps. Thus, our sequential two-stage method not only provides a good estimate of $c$, but also utilizes the advantage of the D-optimal design scheme for choosing design points for the other two parameters with the assistance of a good estimate of $c$. Under a sequential confidence set estimation set up, the proposed method achieves the prescribed estimation accuracy faster than the strict D-optimal design counterpart and the method based on random design as well, which is clearly shown in our numerical study. It is worth noting that the estimate of $c$ can be seriously biased for items with lower discrimination and difficulty levels, and that is why the numerical results for these kinds of parameter combinations are not as stable as others. Hence, some bias correction method for these kinds of items can be applied for further improvement in these cases. The proposed method should be able to be implemented with a standard online testing setup, where examinee's latent trait levels is estimated adaptively. Due to the nature of sequential design, the items to be calibrated can be separately administered to different students and each examinee may be assigned just one or two of them. Thus, we do not have to administer the items to be calibrated to all examinees and the calibration process will not prolong the active test too much. Moreover, please note that the statistical properties here are founded on the martingale theory and the method of stochastic regression that allows the designs to be selected adaptively and sequentially. In theory, the calibration process can even be embedded in different courses of tests. Thus, if we take advantage of modern computing technology fully, the items that cannot be calibrated with sufficient accuracy can be carried over to next active test as long as the data (examinees' abilities and their corresponding responses) can be retained in a database. Of course there are many details to be addressed in a practical situation. However, we believe that the properties of the proposed method also provide useful information for the practical implementation.

## Appendix A: Proofs

First, please note that because the responses of 3PL models are bounded, Condition (C1) – (C3) are easily satisfied as long as the latent trait levels selected for calibration are not equal. Condition (D) is satisfied, for example, a variable length computerized adaptive test is used to estimate the latent trait levels of examinees. It is clear that once Equation 5 is established, then under Condition (C1)-(C3) and (D), the proofs of Theorems 1 and 2 follow easily from arguments similar to those of Chang (2011) and Chang and Lu (2010). So, only highlights of the proofs are given. For the details of proofs, please refer to Chang and Lu (2010).

The proof of the strong consistency of the estimate $\hat{\gamma}$ in Theorem 1 follows by employing the Taylor Expansion Theorem to the log-likelihood function, and then applying the Theorem 7.4.2 of Chow and Teicher (1988); the asymptotic normality of the estimate will follow from a martingale central limit theorem such as Theorem 2.2 of Dvoretzky (1972). Proof of Theorem 2 (i) follows from Chow and Robbins (1965) Lemma 1.

It follows from the definition of $P_i$ that $\partial P_i(\tilde{x}_i)/\partial \beta = (1-c)G_i(\beta)(1-G_i(\beta))\tilde{x}_i$ and $\partial P_i(\tilde{x}_i)/\partial c = 1 - G_i(\beta)$. Let $\sigma_i^2 = P_i(\tilde{x}_i)(1 - P_i(\tilde{x}_i)) > 0$ for all $i \geq 1$, then

$$U_n(\gamma, \tilde{x}, Y) = \sum_{i=1}^{n} \sigma_i^{-2} \left(\frac{\partial P_i}{\partial \beta}', \frac{\partial P_i}{\partial c}\right)' [Y_i - P_i(\tilde{x}_i)] \equiv \sum_{i=1}^{n} z_i.$$

Thus, to prove Theorem 2 (ii) suffices to prove that $\{\sqrt{n}(\hat{\gamma}_n - \gamma_0): n \geq 1\}$ is uniformly continuous in probability (u.c.i.p.; see Woodroofe, 1982, page 10). It is equivalent to show that $\{\sqrt{n}U_n(\gamma_0, x, Y): n \geq 1\}$ is u.c.i.p. Because $U_n(\gamma_0, x, Y) = \sum_{i=1}^{n} z_i$ is a sum of martingale differences. Let $S_n = \sum_{i=1}^{n} z_i$, then for $k, n \geq 1$, we have:

$$\left|\frac{S_{n+k}}{\sqrt{n+k}} - \frac{S_n}{\sqrt{n}}\right| \leq \frac{1}{\sqrt{n}}|S_{n+k} - S_n| + \left[1 - \sqrt{\frac{n}{n+k}}\right]\left|\frac{S_n}{\sqrt{n}}\right|.$$

Hence, the remaining arguments follow from the arguments of Woodroofe (1982) (Example 1.8, page 11) by replacing Kolmogorov's inequality in his arguments with the Hájek-Rényi inequality for martingale differences (Theorem 7.4.8 (iii) of Chow and Teicher, 1988). This implies that $\{\sqrt{n}U_n(\gamma_0, x, Y): n \geq 1\}$ is u.c.i.p., and therefore completes the proof of Theorem 2 (ii).

# References

Ban, J, Hanson, B., Wang, T., Yi, Q. and D. Harris (2001). A Comparative Study of On-line Pretest Item-Calibration/Scaling Methods in Computerized Adaptive Testing, Journal of Educational Measurement.

Carroll, R. J., D. Ruppert, L. A. Stefanski, and C. Crainiceanu (2006). Measurement Error in Nonlinear Models: A Modern Perspective (second ed.), Volume 105 of Monographs on Statistics and Applied Probability. New York: Chapman and Hall/CRC Press.

Chang, Y.-c. I. (2011). Sequential estimation in generalized linear models when covariates are subject to errors. Metrika 73, 93 – 120.

Chang, Y.-c. I. and H.-Y. Lu (2010). Online calibration via variable length computerized adaptive testing. Psychometrika 75(1), 140 – 157.

Chang, Y.-c. I. and Z. Ying (2004). Sequential estimation in variable length computerized adaptive testing. Journal of Statistical Planning and Inference 121(2), 249–264.

Chow, Y. S. and H. Robbins (1965). On the asymptotic theory of fixed-width sequential confidence intervals for the mean. Ann, Math. Statist. 36(2), 457–462.

Chow, Y. S. and H. Teicher (1988). Probability Theory (2ed.). New York, USA: Springer.

Dvoretzky, A. (1972). Asymptotic normality for sums of dependent random variables. In Proc. Sixth Berkeley Symp. Math Statist. Probability, Volume 2, University of California Press, pp. 513–535.

Jones, D. H. and Z. Jin (1994). Optimal sequential designs for on-line item estimation. Psychometrika 59, 59–75.

Kalish, L. A. and J. L. Rosenberger (1978). Optimal Designs for the Estimation of the Logistic Function, Technical Report 33. The Pennsylvania State University, Department of Statistics.

Kingsbury, G. G. (2009). Adaptive item calibration: A process for estimating item parameters within a computerized adaptive test. In D. J. Weiss (Ed.), *Proceedings of the 2009 GMAC Conference on Computerized Adaptive Testing.*

Lai, T. L. and C. Z. Wei (1982). Least Squares Estimates in Stochastic Regression Models with Applications to Identification and Control of Dynamic Systems, Annals of Statist., 10, 154–166.

Makransky, G. (2009). An automatic online calibration design in adaptive testing. In D. J. Weiss (Ed.), *Proceedings of the 2009 GMAC Conference on Computerized Adaptive Testing.*

Patton, J. M. Y. Cheng, K.-H. Yuan and Q. Diao (2013). The Influence of Item Calibration Error on Variable-Length Computerized Adaptive Testing, Applied Psychological Measurement 37, 24-40.

Siegmund, D. (1985). Sequential Analysis: Tests and Confidence Intervals, Springer-Verlag, New York.

Silvey, S. D. (1980). Optimal Design. London: Chapman and Hall.

Van der Linden, W. J. and R. K. Hambleton (1997). Handbook of Modern Item Response Theory. Springer.

Van der Linden, W. J. & Glas, C. A. W. (2000). Cross-validating item parameter estimation in adaptive testing. In A. Boorsma, M. A. J. van Duijn, & T. A. B. Snijders (Eds.), *Essays on item response theory*. New York: Springer.

Wainer, H. (2000). Computerized Adaptive Testing: A Primer (2 ed.), Lawrence Erlbaum Association, Inc., USA.

Woodroofe, M. (1982). Nonlinear renewal theory in sequential analysis. SIAM, Philadelphia, USA.

Table 1: The average estimates, and their corresponding standard deviations and mean square errors of the 2-stage design procedure with $d = 0.5$ and the target coverage probability $= 0.95$ (i.e. $\alpha = 0.05$). For all cases, the true guessing parameter is $c = 0.1$.

| a | b | $\hat{a}$ | $\hat{b}$ | $\hat{c}$ | MSE of $a$ | MSE of $b$ | MSE of $c$ |
|---|---|---|---|---|---|---|---|
| 0.5 | -2 | 0.551(.142) | -1.828(.642) | 0.133(.098) | 0.023 | 0.441 | 0.011 |
| | -1 | 0.525 (.097) | -0.962 (.456) | 0.115 (.057) | 0.010 | 0.209 | 0.003 |
| | 0 | 0.511 (.065) | -0.011 (.358) | 0.102 (.036) | 0.004 | 0.128 | 0.001 |
| | 1 | 0.505 (.051) | 0.976 (.307) | 0.102 (.029) | 0.003 | 0.095 | 0.001 |
| | 2 | 0.501 (.043) | 1.971 (.257) | 0.100 (.022) | 0.002 | 0.067 | <0.001 |
| 1 | -2 | 1.013 (.105) | -1.977 (.164) | 0.106 (.032) | 0.011 | 0.027 | 0.001 |
| | -1 | 1.042 (.149) | -0.969 (.213) | 0.112 (.045) | 0.024 | 0.046 | 0.002 |
| | 0 | 1.014 (.120) | -0.019 (.179) | 0.102 (.030) | 0.015 | 0.032 | 0.001 |
| | 1 | 1.002 (.080) | 0.986 (.123) | 0.100 (.023) | 0.006 | 0.015 | 0.001 |
| | 2 | 1.002 (.056) | 1.990 (.093) | 0.100 (.015) | 0.003 | 0.009 | <0.001 |
| 1.5 | -2 | 1.503 (.102) | -1.998 (.077) | 0.102 (.017) | 0.010 | 0.006 | <0.001 |
| | -1 | 1.517 (.152) | -0.991 (.106) | 0.104 (.027) | 0.023 | 0.011 | 0.001 |
| | 0 | 1.504 (.146) | -0.018 (.128) | 0.101 (.024) | 0.021 | 0.017 | 0.001 |
| | 1 | 1.501 (.095) | 0.995 (.064) | 0.101 (.015) | 0.009 | 0.004 | <0.001 |
| | 2 | 1.499 (.066) | 1.996 (.044) | 0.100 (.010) | 0.004 | 0.002 | <0.001 |
| 2 | -2 | 2.000 (.089) | -2.000 (.034) | 0.100 (.011) | 0.008 | 0.001 | <0.001 |
| | -1 | 2.005 (.150) | -0.999 (.057) | 0.102 (.019) | 0.022 | 0.003 | <0.001 |
| | 0 | 1.995 (.173) | -0.009 (.067) | 0.099 (.021) | 0.030 | 0.005 | <0.001 |
| | 1 | 1.997 (.102) | 0.996 (.040) | 0.100 (.012) | 0.010 | 0.002 | <0.001 |
| | 2 | 1.999 (.067) | 1.998 (.026) | 0.100 (.008) | 0.004 | 0.001 | <0.001 |

Table 2: The average estimates, and their corresponding standard deviation and mean square errors, for the procedure based on a strict D-optimal design with $d = 0.5$ and the target coverage probability $= 0.95$ (i.e. $\alpha = 0.05$). For all cases, the true guessing parameter is $c = 0.1$.

| $a$ | $b$ | $\hat{a}$ | $\hat{b}$ | $\hat{c}$ | MSE of $a$ | MSE of $b$ | MSE of $c$ |
|---|---|---|---|---|---|---|---|
| 0.5 | -2 | 0.639(.234) | **-0.804(.777)** | **0.350(.126)** | 0.074 | 2.032 | 0.078 |
|  | -1 | 0.617(.279) | **-0.254(.769)** | **0.265(.117)** | 0.091 | 1.147 | 0.041 |
|  | 0 | 0.588(.364) | **0.422(.782)** | **0.205(.108)** | 0.140 | 0.789 | 0.023 |
|  | 1 | 0.549(.288) | 1.268(.769) | 0.167(.117) | 0.085 | 0.663 | 0.018 |
|  | 2 | 0.539(.233) | 2.207(.345) | 0.137(.040) | 0.056 | 0.162 | 0.003 |
| 1 | -2 | 1.137(.284) | **-1.549(.687)** | **0.267(.156)** | 0.099 | 0.674 | 0.052 |
|  | -1 | 1.085(.201) | -0.835(.466) | **0.178(.097)** | 0.048 | 0.245 | 0.015 |
|  | 0 | 1.041(.247) | 0.007(.719) | 0.137(.082) | 0.063 | 0.516 | 0.008 |
|  | 1 | 1.025(.390) | 1.089(.580) | 0.123(.071) | 0.153 | 0.344 | 0.006 |
|  | 2 | 0.999(.114) | 1.994(.366) | 0.104(.044) | 0.013 | 0.134 | 0.002 |
| 1.5 | -2 | 1.549(.282) | -1.851(.463) | **0.203(.131)** | 0.082 | 0.236 | 0.028 |
|  | -1 | 1.545(.202) | -0.942(.300) | 0.136(.067) | 0.043 | 0.093 | 0.006 |
|  | 0 | 1.508(.235) | -0.032(.464) | 0.112(.068) | 0.055 | 0.216 | 0.005 |
|  | 1 | 1.503(.212) | 1.023(.353) | 0.108(.062) | 0.045 | 0.125 | 0.004 |
|  | 2 | 1.505(.134) | 2.007(.134) | 0.102(.028) | 0.018 | 0.018 | 0.001 |
| 2 | -2 | 1.976(.332) | -1.834(.589) | **0.197(.142)** | 0.111 | 0.374 | 0.029 |
|  | -1 | 2.012(.233) | -0.991(.242) | 0.117(.062) | 0.054 | 0.059 | 0.004 |
|  | 0 | 1.997(.220) | -0.013(.171) | 0.102(.054) | 0.049 | 0.029 | 0.003 |
|  | 1 | 1.988(.215) | 1.022(.288) | 0.105(.039) | 0.046 | 0.084 | 0.002 |
|  | 2 | 1.912(.373) | 2.117(.490) | 0.116(.094) | 0.146 | 0.253 | 0.009 |

Table 3: Sample sizes and coverage probabilities for the 2-stage design and the strict D-optimal design procedures with d = 0.5 and the target coverage probability = 0.95. The true guessing parameter c = 0.1 for all cases.

| | | 2-stage design | | | | Strict D-optimal | | | |
|---|---|---|---|---|---|---|---|---|---|
| | | | Coverage prob. of | | | | Coverage prob. of | | |
| a | b | n | a | b | c | n | a | b | c |
| 0.5 | -2 | 524.42(54.12) | 0.999 | 0.995 | 1.000 | 1432.48(3322.87) | 0.986 | 0.997 | 1.000 |
| | -1 | 562.37(72.97) | 1.000 | 1.000 | 1.000 | 2853.71(4819.17) | 0.980 | 0.980 | 1.000 |
| | 0 | 742.04(123.97) | 1.000 | 1.000 | 1.000 | 4508.72(3665.53) | 0.968 | 0.907 | 1.000 |
| | 1 | 1083.62(197.17) | 1.000 | 1.000 | 1.000 | 4348.04(3179.25) | 0.986 | 0.977 | 0.997 |
| | 2 | 1555.37(241.51) | 1.000 | 1.000 | 1.000 | 3785.31(3639.98) | 0.995 | 0.994 | 1.000 |
| 1 | -2 | 1211.48(200.99) | 1.000 | 1.000 | 1.000 | 1188.60(2717.00) | 0.986 | 0.985 | 1.000 |
| | -1 | 635.42 (69.72) | 1.000 | 1.000 | 1.000 | 1440.42(3814.30) | 0.999 | 0.993 | 1.000 |
| | 0 | 890.06(132.47) | 1.000 | 1.000 | 1.000 | 1882.64(2804.74) | 0.986 | 0.934 | 1.000 |
| | 1 | 1755.53(263.51) | 1.000 | 1.000 | 1.000 | 4039.29(2132.58) | 0.988 | 0.963 | 1.000 |
| | 2 | 3175.46(407.92) | 1.000 | 1.000 | 1.000 | 5085.21(4041.70) | 1.000 | 0.999 | 1.000 |
| 1.5 | -2 | 2745.17(356.14) | 0.999 | 1.000 | 1.000 | 1555.80 (553.21) | 0.995 | 0.995 | 1.000 |
| | -1 | 1147.28(205.08) | 1.000 | 1.000 | 1.000 | 1008.93(590.93) | 0.998 | 0.995 | 1.000 |
| | 0 | 1099.61(180.03) | 0.999 | 1.000 | 1.000 | 1687.20(1099.22) | 0.987 | 0.974 | 1.000 |
| | 1 | 2684.63(318.50) | 1.000 | 1.000 | 1.000 | 6286.53(1462.70) | 0.993 | 0.993 | 1.000 |
| | 2 | 5513.54(529.82) | 1.000 | 1.000 | 1.000 | 7123.90(2662.86) | 1.000 | 1.000 | 1.000 |
| 2 | -2 | 5023.19(485.25) | 1.000 | 1.000 | 1.000 | 3152.87(1157.12) | 0.968 | 0.976 | 1.000 |
| | -1 | 2022.2 (298.34) | 1.000 | 1.000 | 1.000 | 1780.11 (353.23) | 0.990 | 0.997 | 1.000 |
| | 0 | 1535.33(260.81) | 0.999 | 1.000 | 1.000 | 1857.59 (571.81) | 0.990 | 0.995 | 1.000 |
| | 1 | 3933.47(413.99) | 1.000 | 1.000 | 1.000 | 5749.74(1988.83) | 0.992 | 0.994 | 1.000 |
| | 2 | 8682.23(684.42) | 1.000 | 1.000 | 1.000 | 10753.10(1771.50) | 0.939 | 0.994 | 1.000 |

Table 4: The average estimates, and their corresponding standard deviation and mean square errors, for the procedure based on a random design using a normally distributed latent trait levels of examinees with $d = 0.5$ and the target coverage probability $= 0.95$ (i.e. $\alpha = 0.05$). For all cases, the true guessing parameter is $c = 0.1$.

| a | b | $\hat{a}$ | $\hat{b}$ | $\hat{c}$ | n |
|---|---|---|---|---|---|
| 0.5 | -2 | **0.685(.172)** | **-0.346(.821)** | **0.401(.118)** | 4391.78(7393.41) |
|  | -1 | **0.673(.279)** | **-0.162(.623)** | **0.315(.100)** | 2104.16(5646.16) |
|  | 0 | **0.669(.122)** | **0.888(.534)** | **0.264(.083)** | 2344.70(3255.73) |
|  | 1 | **0.627(.102)** | **1.463(.342)** | **0.195(.057)** | 3067.64(2456.64) |
|  | 2 | **0.591(.111)** | **2.193(.230)** | **0.151(.033)** | 4151.27(2756.12) |
| 1 | -2 | 1.113(.256) | **-0.915(.910)** | **0.403(.162)** | 660.59(296.49) |
|  | -1 | 1.110(.171) | **-0.552(.540)** | **0.229(.101)** | 799.55(250.76) |
|  | 0 | 1.055(.125) | 0.102(.317) | 0.135(.054) | 1540.49(1269.78) |
|  | 1 | 1.026(.085) | 1.012(.094) | 0.106(.023) | 2428.43(793.34) |
|  | 2 | 1.005(.072) | 1.994(.075) | 0.100(.012) | 4694.45(673.36) |
| 1.5 | -2 | **1.390(.372)** | **-1.191(.116)** | **0.336(.240)** | 1311.86(622.70) |
|  | -1 | 1.539(.167) | -0.929(.265) | 0.125(.066) | 873.65(156.60) |
|  | 0 | 1.512(.141) | 0.000(.140) | 0.100(.030) | 1195.49(349.92) |
|  | 1 | 1.501(.104) | 0.995(.054) | 0.099(.014) | 2589.11(445.69) |
|  | 2 | 1.496(.134) | 1.998(.042) | 0.099(.007) | 7351.16(1074.22) |
| 2 | -2 | **1.789(.487)** | **-1.507(.987)** | **0.242(.241)** | 3309.08(1584.42) |
|  | -1 | 1.991(.199) | -0.975(.242) | 0.108(.061) | 1497.44(266.96) |
|  | 0 | 1.987(.166) | -0.006(.102) | 0.097(.024) | 1419.92(286.99) |
|  | 1 | 1.994(.113) | 0.998(.037) | 0.099(.010) | 3651.53(433.01) |
|  | 2 | 1.995(.092) | 2.000(.031) | 0.100(.005) | 12852.29(1260.36) |

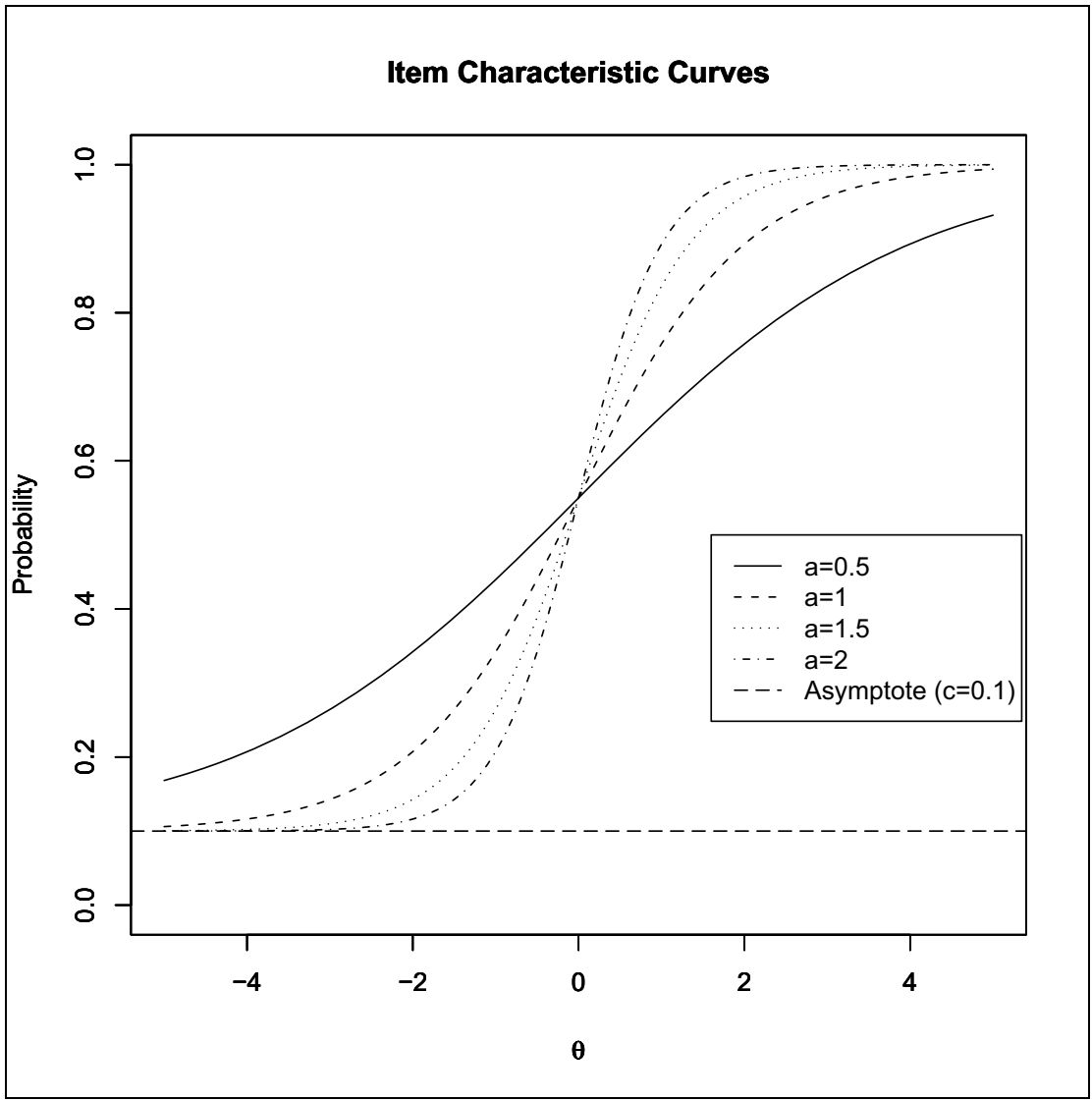

Figure 1: Item characteristic curves for items with $c = 0.1$, $b = 0$, different values of a's and their asymptote.

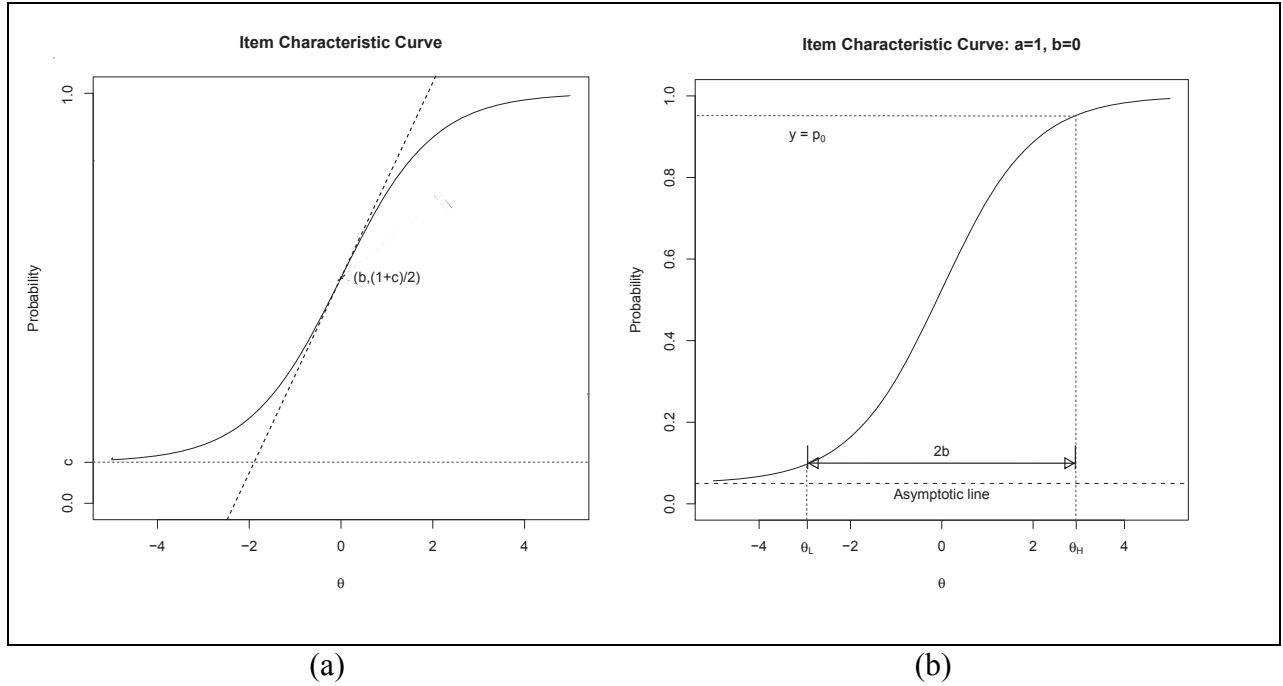

(a)          (b)

Figure 2: The property of reflectional symmetry (a) and the threshold of lower abilities, say, $\theta_L$ with estimates $a$ and $b$ (b).

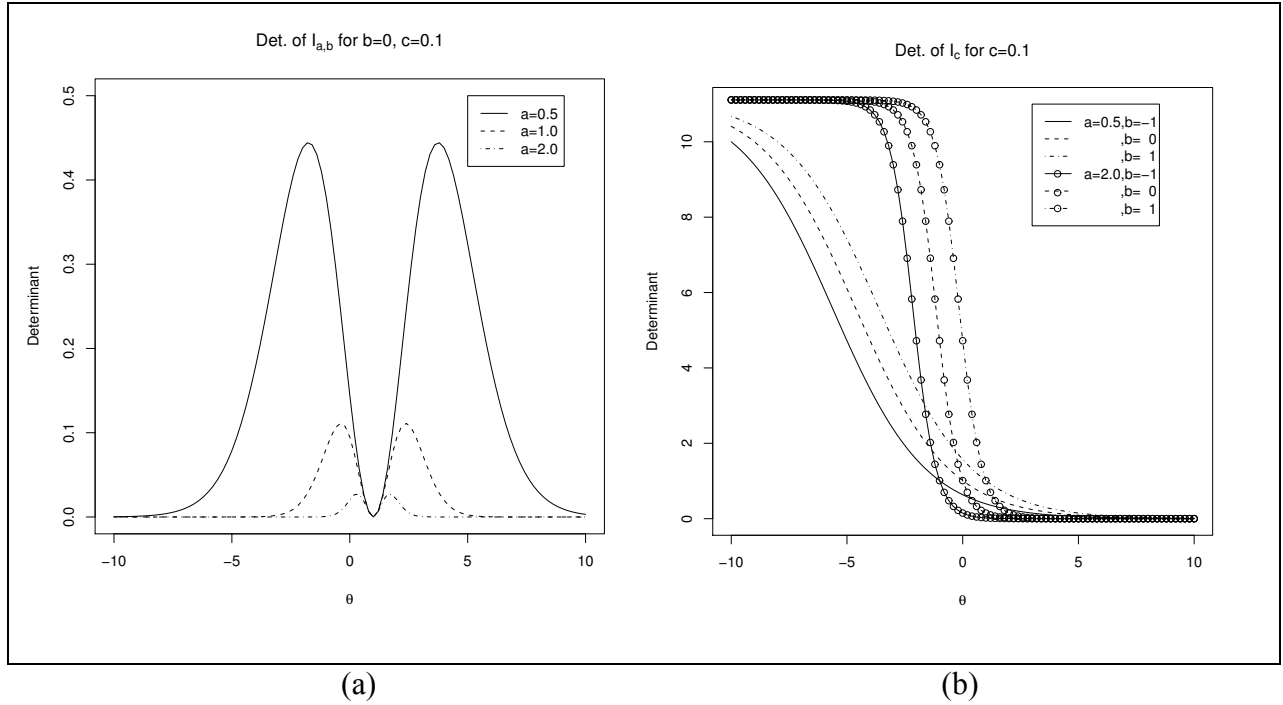

(a) (b)

Figure 3: Information curves of parameters ($a$, $b$) and $c$ as functions of $\theta$.